\documentclass[apj]{emulateapj}

\def\gtorder{\mathrel{\raise.3ex\hbox{$>$}\mkern-14mu
             \lower0.6ex\hbox{$\sim$}}}
\def\ltorder{\mathrel{\raise.3ex\hbox{$<$}\mkern-14mu
             \lower0.6ex\hbox{$\sim$}}}

\begin{document}

\title{A Survey About Nothing: Monitoring a Million Supergiants for Failed Supernovae}

\author{Christopher S. Kochanek\altaffilmark{1,3}, 
John F. Beacom\altaffilmark{1,2,3}, 
Matthew D. Kistler\altaffilmark{2,3}, Jos{\'e} L. Prieto\altaffilmark{1,3}, 
Krzysztof Z. Stanek\altaffilmark{1,3}, Todd
A. Thompson\altaffilmark{1,3}, and Hasan Y{\"u}ksel\altaffilmark{2,3}}

\altaffiltext{1}{Dept.\ of Astronomy, The Ohio State University, 140 W.\ 18th Ave., Columbus, OH 43210}
\altaffiltext{2}{Dept.\ of Physics, The Ohio State University, 191 W.\ Woodruff Ave., Columbus, OH 43210}
\altaffiltext{3}{Center for Cosmology and AstroParticle Physics, The Ohio State University, 
191 W.\ Woodruff Ave., Columbus, OH 43210}

\shorttitle{Do Supergiants Vanish?}

\begin{abstract}

Extragalactic transient searches have historically been limited to
looking for the {\it appearance} of new sources such as supernovae.
It is now possible to carry out a new kind of survey that will do the
opposite, that is, search for the {\it disappearance} of massive
stars.  This will entail the systematic observation of galaxies within
a distance of 10 Mpc in order to watch $\sim 10^6$ supergiants.
Reaching this critical number ensures that {\it something} will
occur yearly, since these massive stars must end their lives with a
core collapse within $\sim 10^6$ years.  Using deep imaging and
image subtraction it is possible to determine the fates of these stars
whether they end with a bang (supernova) or a whimper (fall out of
sight).  Such a survey would place completely new limits on the total
rate of all core collapses, which is critical for determining the
validity of supernova models. It would also determine the properties of supernova
progenitors, better characterize poorly understood optical transients,
such as $\eta$ Carina-like mass ejections, find and characterize large
numbers of Cepheids, luminous blue variables and eclipsing binaries, and allow the
discovery of any new phenomena that inhabit this relatively unexplored parameter
space.

\end{abstract}



\section{Introduction}
\label{sec:int}

In general, it is easier to notice a phenomenon by presence,
rather than absence. But absence can be the crucial clue,
as in the case of the dog that didn't bark in the night 
(Doyle 1892).  This is also true in
astronomy.  For example, while the brightness of
supernovae (SNe) enabled their study by naked-eye astronomers, it has
only recently been possible to detect the progenitors of such events
-- although to date almost all were found through serendipity rather
than careful planning. With modern 8-m class
telescopes, wide field cameras, and image subtraction, it is now
possible to conduct a comprehensive survey of massive stars and
determine all causes of death.  While most will probably
die as a bright, core-collapse SN, it is likely that some
fraction do not follow this route, producing either an
exceedingly-dim SN or else completely ``fail'' and collapse
directly to a black hole (BH) with no optical fireworks at all.
Little is currently known about the optical 
signatures of BH formation even though we believe it is the
typical end of the most massive stars ($M\gtorder 25 M_\odot$) and
could be a common end at lower masses ($8\ltorder M \ltorder 25M_\odot$) 
given the theoretical challenges to producing successful SN explosions.

Consider a survey that will watch enough supergiants before they
suffer a core collapse to make a quantitative study of their final
states.  Since the remaining lifetime of a star that has reached this
phase is $\sim 10^6$~yr, this requires observing at least 
$\sim 10^6$ supergiants to expect an appreciable number of events over the
duration of the survey.  Equivalently, one must survey enough galaxies 
to observe $\sim 1$~SN/year. This number can be attained by monitoring only
$\simeq 30$ galaxies within 10~Mpc, as we explain below.  While this
is a challenging observational project, in the long run there is far
more physical information in determining the fates of individual stars
of various types than there is in inferring their fates based on the
mean properties of their host galaxies.  Even modest observing efforts
over the next five years could either find examples of failed SNe or
limit their rates to be significantly below those of normal SNe, with
important consequences for both supernova physics and efforts to
detect gravitational waves.

Importantly, a survey designed to detect disappearance will necessarily also
be excellent for appearance studies with guaranteed results on known
phenomena such as normal SNe, heavily obscured SNe, $\eta$ Carina-like 
outbursts, eclipsing binaries, novae, luminous blue variables (LBVs) 
and Cepheid variables. These are in turn important for the late phases 
of massive star evolution, formation rates of binaries with compact 
objects, total SN rates and the local distance scale (binaries and 
Cepheids).  Beyond these certainties,
new classes of events surely await discovery. 


\section{Autopsies of Massive Stars}
\label{sec:theend}

\begin{figure}[t]
\plotone{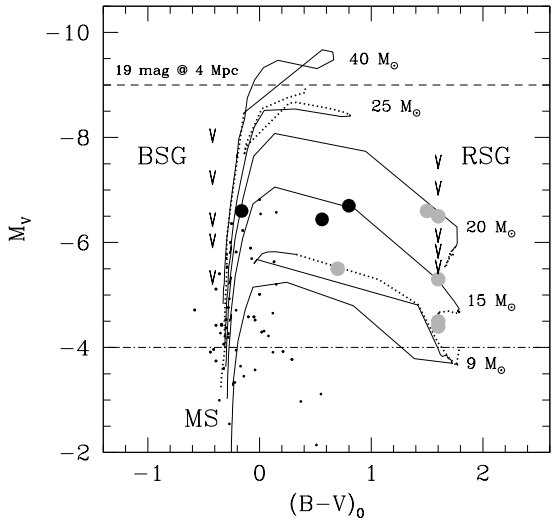} 
\caption{A color-magnitude diagram showing evolutionary tracks
(Lejeune \& Schaerer 2001) for various masses of progenitors at solar
metallicity, with the star's last $5 \times 10^5$ years as a dotted line.  The
labels approximately mark the locations of the main sequence, blue
supergiants and red supergiants.  The black circles are 
progenitors in pre-explosion images (1987A the bluest, 2004et and
1993J in the yellow range) with measured B--V colors, the gray
circles are progenitors without measured B--V colors, and the
arrows are upper limits.  The points are Wolf-Rayet stars in the Magellanic
Clouds (Massey 2002).  The horizontal dashed lines mark the typical
depth of a SNe survey and the depth required for survey for failed SNe.
For the gray points we estimated the B--V color from
either the I magnitude or the measured V--I color assuming 
the progenitors were K5 supergiants.  For the upper limits we
used the color of a K5 supergiant for the Type II SN and
a fixed blue color for the Type Ib/c SN.
The SN progenitors are taken from the tabulation
in Li et al. (2007) and references therein.
}
\label{fig:cmd}
\end{figure}

\begin{figure}[t]
\plotone{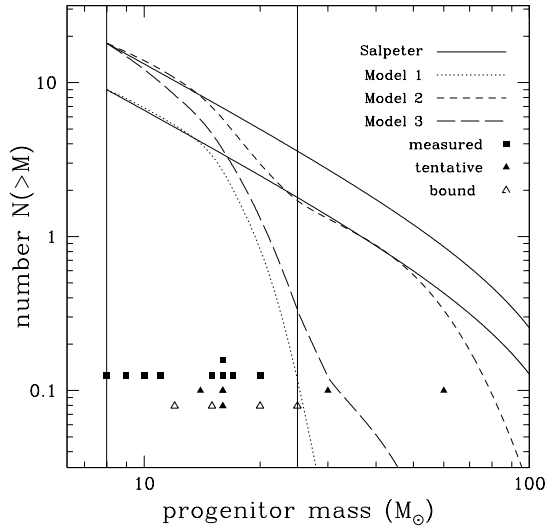}
\caption{Integral progenitor mass distributions $N(>M)$. The filled squares,
  filled triangles, and open triangles show the well-estimated
  progenitor masses, possible masses, and upper bounds
  on the mass from Li et al.~(2007), respectively.  We consider
  3 models for $N(>M)$ as described in the text.  Model
  1 is based on only the 9 measured progenitor masses.  Models 2 and
  3 include all 18 estimates and treat the less 5 reliable tentative
  measurements as either measurements (model 2) or upper bounds
  (model 3). The Salpeter models (solid lines) are normalized to either 9
  SNe for comparison to Model 1 or to 18 SNe for comparison
  to Models 2 and 3.  The
  vertical lines demarcate the canonical successful SN mass
  range of $8\ltorder M\ltorder25$\,M$_\odot$. In all three
  models, the number of observed high mass progenitors is less
  than expected. }
\label{fig:prog}
\end{figure}

Searches for SN explosions have exploded (as it were) in the past
decade, both locally and at cosmological distances (e.g., 
Evans 1997;
Li et al. 2000; 
Riess et al. 2004;
Astier et al. 2006; 
Miknaitis et al. 2007; 
Frieman et al. 2008).  Finding SNe is relatively easy because
of their enormous peak brightness.  Identification of SN progenitors
has lagged because there are few SN host galaxies that, by
coincidence, had the required very deep pre-SN images.  The existing
samples of progenitors (e.g., Smartt et al.~2004, Li et al.~2007) are
dominated by red supergiants with estimated initial masses in the
expected range ($8 M_\odot \ltorder M \ltorder 25 M_\odot$), with at
least hints of a dearth of more massive progenitors (see
Figs.~\ref{fig:cmd} and \ref{fig:prog}).  We quantify this by comparing
the integral distribution of progenitor masses from  Li et al. (2007) 
with the distribution from a Salpeter IMF for $8M_\odot < M < 150 M_\odot$.
We neglect 2000ew (which is simply called ``low mass") and 2000ds (whose mass limit of
$<7M_\odot$ is below our $8M_\odot$ cutoff); including them would only
strengthen the argument that follows. We divided the remaining
progenitors into three groups: 9 systems with mass estimates,
5 (2005gl, 2004dj, 1999gi 2001du and 2004gt) with possible masses,
and 4 (1999em, 1999an, 1999br, 2001B) with only upper limits on
the progenitor mass.

We estimate the differential mass distribution of progenitors $dN/dM$
as follows.  For each progenitor we have a prior on its mass, the 
Salpeter IMF $P_S(M)\propto M^{-2.35}$ with $8 M_\odot \leq M \leq 150 M_\odot$, 
and then a probability distribution $P_i(M_i|M)$ relating the mass $M$ and the 
mass estimates $M_i$ from Li et al.~(2007).  Combining the two, the Bayesian
estimate for the progenitor mass is the product $P_i(M|M_i)\propto P_S(M) P_i(M_i|M)$ 
normalized to unity.  If a progenitor has a mass estimate, we use a
log-normal probability distribution for $P_i(M_i|M)$ based on the reported
uncertainties. If a progenitor only has a mass limit $M_i$, we use a flat
probability distribution extending from $8M_\odot$ to the mass limit $M_i$
for $P_i(M_i|M)$.  The cumulative mass function of progenitors, 
\begin{equation} N(>M) = \int_M^{150M_\odot} \sum_{i=1}^N P_i(M|M_i) dM,
\end{equation}
is simply the mass integral of the sum of the $i=1\cdots N$ progenitor
probability distributions $P_i(M|M_i)$.  

Figure~\ref{fig:prog} presents three different estimates for $N(>M)$ 
from the data and compares these estimates to the expectation for a
Salpeter IMF.  In Model 1 we use only the $N=9$ progenitors with 
mass estimates.  In Model 2 we use all $N=18$ systems, adding
the 5 with tentative mass estimates as measurements.  In Model 3 we 
use all $N=18$ systems, but treat the 5 tentative mass estimates as 
upper bounds rather than measurements.  The results certainly suggest
a deficit of high mass progenitors, but at low statistical significance.
In a sample of 9 (18) progenitors we would expect 1.8 (3.6) more massive
than $25M_\odot$, while the current samples include only 0.5 (2), where
the details clearly depend on how we build the distributions.
It is more difficult to address whether the apparent
deficit is simply a selection effect.  The more massive progenitors 
should not be intrinsically fainter in the visual bands, even though the
bulk of their emission is at shorter wavelengths, but systematic effects
such as a correlation between age and extinction could easily produce
a similar deficit.  For example, Prieto et al. (2008b) found that
the progenitor of SN2008S (which may be an LBV eruption rather than a
SN) is so enshrouded in dust that it was only visible from dust 
emission in the mid-IR. 

Moreover, optical searches for SNe and their progenitors provide little
direct information on the intriguing question of whether there are massive
stars that end their lives by forming BHs without the dramatic
visual signature of an explosion.
The upper bound on the potential rate of failed SNe is roughly equal
to the rate of successful SN.  First, the concordance of massive star
formation rates and SNe rates and the non-detection of a diffuse
SN neutrino background both indicate that the rate of failed SNe
cannot significantly exceed the rate of observed SN (Hopkins \& Beacom
2006).  Second, the non-observation of any neutrino bursts over the
last twenty-five years (Beacom et al.~2001, Alekseev \& Alekseeva 2002,
Ikeda et al. 2007) sets a weak upper bound on all core
collapses in the Galaxy of $\ltorder 12$ events per
century (95\% confidence) as compared to the rate of roughly 1 SN per century (e.g.,
van den Bergh \& Tammann 1991; Cappellaro et al. 1999).  

A crude lower bound may be obtained from the (albeit poorly constrained)
formation rate of BHs by all possible mechanisms.  A population
census of BHs and neutron stars (excepting pulsars) in the Galaxy is presently
impractical because they can only be found through special populations
of binaries such as active X-ray binaries or astrometric binaries
(Gould \& Salim 2002), or through the modest contribution of these
compact objects to Galactic microlensing rates (Gould 2000).  The
existence of high-mass X-ray binaries (HMXBs), in which we observe a
massive BH orbiting a short-lived massive star (like the
spectacular system M33-X7 with a $16M_\odot$ BH orbiting a
$70M_\odot$ star (Orosz et al. 2007)), means that the present day
BH formation rate is non-zero (Bethe et al. 2007).  The observed pattern of
stellar element abundances may require that most stars more
massive than $\simeq 25 M_\odot$ collapse to form BHs in order
to avoid overproducing heavy elements (Heger et al. 2003).\footnote{There
is the intriguing observation by Muno et al. (2006) of a 
probable magnetar in a star cluster containing $M\simeq 35M_\odot$ stars,
but Belczynski \& Taam (2008) recently presented a binary evolution scenario 
in which Roche lobe overflow allows some 50--80$M_\odot$ stars
to form neutron stars.}
For a Salpeter IMF in which stars with $8M_\odot \ltorder M \ltorder 25 M_\odot$
become neutron stars after a classical SN and higher mass stars
become BHs, the BH formation rate is $\sim 25\%$
that of normal SNe. Such simple estimates are consistent with the 
simulations of Zhang et al. (2007).  

\begin{figure}[t]
\plotone{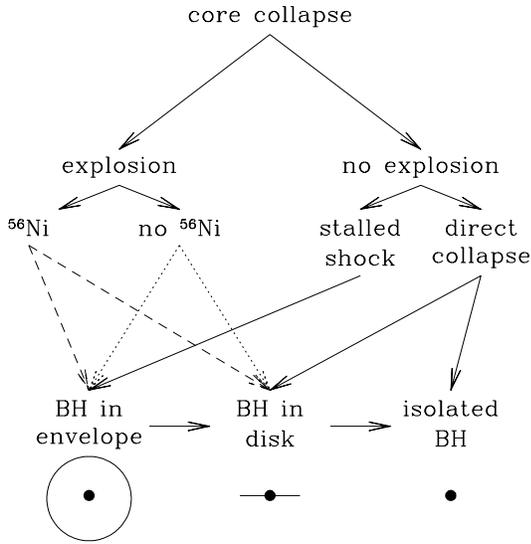}
\caption{Possible outcomes in forming a black hole.  
  The optical signatures of the ``no explosion'' scenarios are
  little explored.
  }
\label{fig:outcomes}
\end{figure}

Beyond these semi-empirical limits we must rely on theoretical studies
of core collapse.  
Despite intense theoretical and computational efforts 
(in 1D, e.g., Rampp \& Janka 2000;
Liebend\"orfer et al.~2001;
Thompson et al.~2003;
Sumiyoshi et al.~2005;
in 2D:
e.g., Fryer 1999;
Buras et al.~2003, 2006;
Livne et al.~2004;
Ohnishi et al.~2006;
and in 3D:
e.g., Fryer \& Warren 2002) 
it is difficult to simulate the evolution of any star with realism,
and most attempts to produce SNe fail.
The steady increase in sophistication of the models over the last
four decades has not unambiguously reduced the difficulties,
although recently a 2D calculation of a relatively low-mass progenitor
(11.2\,M$_\odot$) led to a weak neutrino-driven explosion
(Buras et al.~2006) and there may be new mechanisms associated
with less restrictive simulation geometries 
(Blondin et al.~2003; Burrows et al.~2006; Scheck et al.~2008).
What is particularly worrisome in the history of simulating core 
collapse is that the theoretical effort is focused on making SNe 
succeed (particularly for $M\approx10$\,M$_\odot$) because they are
observed, without significant constraint on whether Nature is 
any more successful at producing explosions than theorists 
(Gould \& Salim 2002).  As we have
reviewed above, there are no observational constraints barring $\sim 50\%$
of  $8M_\odot \ltorder M \ltorder 25 M_\odot$ stars from forming BHs.
In short, the rate of BH formation
could well be comparable to the rate of normal SNe even for relatively
low-mass progenitors.

The rate of failed SNe depends on the optical signatures of 
BH formation, and, unfortunately, we lack a clear prediction of
these signatures.  The possibilities literally range from a nearly
normal SN to the star simply vanishing.
The observed masses of BHs (e.g., Casares
2006) show a distinct gap between neutron stars ($M_{NS}\simeq
1.4M_\odot$) and BHs ($M_{BH} \gtorder 4$), so one would expect
a significant difference in the external signatures rather than a
simple continuum of properties. 
Figure~\ref{fig:outcomes} sketches possible outcomes.  

One scenario entails a successful shock leading to a visible explosion, 
with collapse to a BH after material falls back onto the neutron 
star (e.g., Woosley \& Weaver 1995).  There are arguments from studies of 
the early-time accretion from the envelope (e.g., Chevalier 1993; 
Fryer et al.~1996) supporting this route.  Balberg et al. (2000, 2001),
building on Zampieri et al (1998), 
considered the visible signature from accretion onto a BH following
a SN in some detail and found that the detectability of the accreting BH
depended critically on the ejected mass of radioactive elements.
The accretion luminosity starts (essentially) Eddington-limited
and then decreases as a power law, while the radioactive decay-powered luminosity
is initially far brighter and decreases exponentially.  The time at
which accretion dominates is determined by the mass of the ejected
radioactive elements, particularly the longer-lived $^{44}$Ti. 
For a normal SN ($\sim 0.05M_\odot$ of $^{56}$Ni)
the time scale is very long ($\sim 10^3$ years), while for   
a very low energy SN like 1997D ($\sim 10^{-3}M_\odot$ of $^{56}$Ni) it 
can be very short ($\sim$ years).  In all cases, the BH
is very faint when it emerges.  An alternative is the ``collapsar''
model with a $\gamma$-ray burst and an optical afterglow
superposed on a SN (e.g., MacFadyen \& Woosley 1999).   These
cannot be the dominant signature of BH formation since
they are too rare and are likely confined to metal-poor galaxies
(Stanek et al. 2006).  

The second possibility is that a shock either never forms or stalls
before reaching the stellar surface (``prompt'' formation, Heger et al. 2003).  
Given the challenges in producing successful explosions, this could
well be a common outcome at all masses.  The simple ``direct collapse'' 
scenario may be possible if the core collapses directly into
a BH with the envelope simply following it in 
afterward, as seen in some of the simulations of Duez et al. (2004).
At higher progenitor rotation rates, Duez et al. (2004) found that there
can be a residual accretion disk around the BH.
Alternatively, the core collapses to form a neutron star
with a stalled shock, followed by an accretion 
induced collapse of the neutron star to form a BH.
Most studies of this scenario have focused on the collapse
and its neutrino signature without examining the fate of
the remainder of the stellar envelope (e.g., 
Baumgarte et al. 1996;
Liebend\"orfer et al. 2004; 
Sumiyoshi et al.  2007). We are aware of
no studies of the expected optical signatures for
these scenarios.  

Given an ill-constrained rate, only partially explored pathways, and
poorly constrained optical signatures, the safest way to proceed 
is through observations:  monitor a large enough sample of massive
stars sufficiently deeply to detect all possible outcomes from a 
classical SN to a direct collapse with no signature other than 
dissapearance.

\section{How to Write Obituaries For Massive Stars }
\label{sec:survey}

The ultimate objective is to monitor the health of a sufficient number of
massive stars to directly measure their death rates as a function of
luminosity, temperature, and metallicity.  There are
three distinct challenges in this program: building catalogs of massive
stars to observe, recognizing the death of a star, and observing 
enough stars to have an interesting event rate.  

Cataloging is probably the most difficult problem. Although it does
not impact determining the relative rates of normal and failed SN, it is 
important for determining absolute rates.  
With 8-m telescopes or the {\it Hubble Space Telescope (HST)} it is feasible to regularly measure the flux 
of a $10 M_\odot$ supergiant with $M_V \simeq -4$~mag (see Fig.~\ref{fig:cmd}).  
With no extinction this 
corresponds to $V\simeq 26$~mag (25~mag) at a distance of $D=10$~Mpc (6~Mpc) 
and requires typical exposure times for a signal-to-noise ratio of 10 
and neglecting the diffuse emission from the galaxy of 60 (6) minutes for a
seeing-limited 8-m telescope.  For catalogs, the problem is the
blending of the stars, since the physical resolution of a ground-based
telescope at the galaxy is $5 (D/\hbox{Mpc})$~pc and the massive stars 
tend to be clustered.  Ideally, a single  epoch
of observations with {\it HST} would greatly simplify producing catalogs
while simultaneously providing the flux calibrations needed to use
Cepheids or eclipsing binaries to constrain the distance ladder.
  
Recognizing the death of a star is much easier because
image subtraction (e.g., Alard \& Lupton 1998)
can determine the fate of a star even if it was confused with
other stars in the initial accounting.  In all scenarios, the
final state is the {\it absence} of the star, so a robust signature
of a star's death would be that its flux disappears and does not reappear.  
This is more easily done for failed SN with a minimal optical transient 
at death because, for normal SN, it may take many years
for the dying star to fade to be significantly less luminous 
than before it died.  Other known
sources (see \S\ref{sec:science}) either appear before disappearing
(e.g., novae), vary (ir)regularly (Cepheids, LBVs, eclipsing binaries)
or reappear after disappearing (R Coronae Borealis (RCB) stars) on reasonable time scales.
As we will demonstrate in \S4, the rate of false positives is easily
managed.

Finally, the rate of normal core-collapse SNe in the target sample sets a crucial
scale for the feasibility of such a survey.  The sample must produce
roughly one normal SN/year in order for a limit on failed SN
to be significant.  Since one is limited by technology to nearby galaxies, 
we start with the Karachentsev et al. (2004) catalog of neighboring galaxies, which is
designed to be $\sim 80\%$ complete to a distance of 8~Mpc.  We
estimate the relative core collapse SN rate of the galaxies using the
results of Cappellaro et al. (1999) and then normalize the total rate
to match that observed for these galaxies from 1970-2007 based on the
Sternberg Astronomical Institute SN catalogs (see Ando et al. 2005).  The resulting
predicted and observed rates for the individual galaxies agree well,
although the absolute, total normalization ranges from $0.56$/year for
SNe from 1970-2007 to $1.1$/year if we restrict ourselves to the
``modern'' era of robotic surveys (1997-2007).  We lose 10\% of the
expected rate by eliminating highly inclined galaxies (axis ratios
$<0.3$). Only 40 galaxies need to be observed (30 Northern with
$\hbox{Dec}>-10^\circ$)\footnote{In rough order of increasing observational
cost per SN, they are M101, M81, NGC5194, (NGC5236), NGC2403,
(NGC4594), M82, NGC6946, NGC4258, NGC4736, NGC4826, (NGC1313), IC342,
NGC2903, (NGC7793), (NGC3621), NGC3627, (NGC247), (NGC300), NGC4236,
NGC925, NGC4449, NGC628, (NGC5068), NGC3368, M31, NGC4395, NGC3077,
NGC4605, NGC4214, NGC3351, NGC3344, NGC6503, M33, (NGC5253), IC2574,
NGC672, NGC5474, NGC3489 and (NGC5102). The parentheses indicate
Southern galaxies ($\hbox{Dec}<-10^\circ$).  The observational cost
includes the effects of distance and that M31 and M33 require multiple
pointings for a typical $0.25$~sq.~deg. camera.  We have not corrected
for Galactic extinction, and note that the levels for IC342 ($A_B
\simeq 2.4$) and NGC6946 ($A_B\simeq 1.5$) are uncomfortably high.
We also note, however, that NGC6946 has had 9 (!) SNe over the last 
century, far more than any other galaxy on this list.
 } to
cover 90\% of the expected rate.

\begin{figure}[t]
\plotone{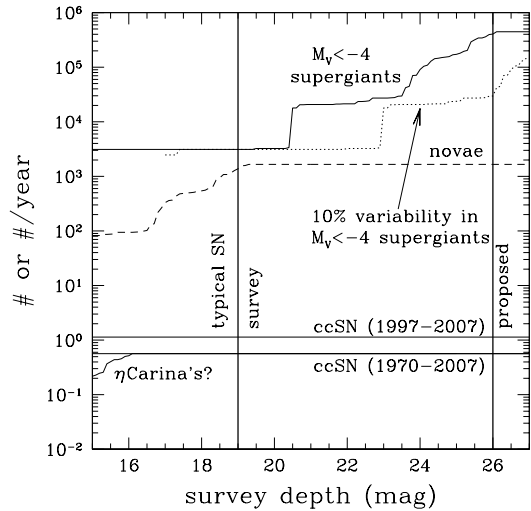}
\caption{Numbers or rates for ``expected'' sources among the
Karachentsev et al. (2004) catalog of nearby Galaxies as a function of
survey depth.  Explosive, luminous events such as SNe, $\eta$
Carina-like outbursts and novae are relatively easy to detect even at
the $19$~mag depth of a typical local SN survey.  Monitoring a
sufficient number of massive stars or the variability of those massive
stars requires far deeper observations.  The conspicuous jumps in the
numbers occur where observing the phenomenon in the SMC/LMC, M31/M33
and the M81 group becomes possible.  The heavy vertical lines mark
the depth of a typical SN survey and the depth required for the most
distant galaxies in our sample.}
\label{fig:rate}
\end{figure}


For these galaxies, the estimated core collapse SN rates are 0.46--0.90 
per year depending on whether we use the lower 1970-2007 or the higher
1997-2007 rate normalizations.  A survey restricted to the Northern
galaxy  sample would have modestly lower rates of 0.35--0.71 per year.
We suspect, and can argue statistically at roughly 90\% confidence,
that the higher normalization of the last decade is correct, where
the change in efficiency is presumably due to the introduction of automated surveys 
(e.g., KAIT, Li et al. 2000), enormous improvements in the equipment 
available to amateurs, and greater community interest in SNe.  
In any case, this sample of galaxies has a high enough SN rate that
a failure to find candidate failed SNe over a period of 5 years
sets an interesting limit.

An alternate way of considering the question is that Hartman et
al. (2006) found approximately 1400 $M_V\ltorder -5$ supergiants in their survey of
M33.  Using the same scalings as for the SN rates, approximately
correcting to $M_V<-4$, and normalizing by
the luminosity of M33, our neighboring galaxy sample contains
approximately $\sim 10^6$ supergiant stars, and so with mean lifetimes of
$\sim 10^6$ years one should see $\sim 1.0$ SNe per year.
Figure~\ref{fig:rate} illustrates the visibility of SNe, $\eta$
Carina-like outbursts, novae\footnote{Normalized using the Darnley et
al. (2006) rates for M31 and assuming all novae peak at
$M_V=-11$.}, $M_V<-4$ supergiants and 10\% variability in such
stars as a function of survey depth.


\section{Guaranteed science}
\label{sec:science}

The full yield of a survey for failed SNe will depend on the monitoring cadence that
the program can sustain.  Several SNe should be found in the galaxies 
under watch, so that deep pre-explosion images will be available to identify 
their progenitors.  In some respects, this is similar to other attempts to
survey nearby galaxies for later identification of SN progenitors 
(e.g., Crockett et al. 2007).  It differs significantly from these programs 
in emphasizing monitoring and image subtraction rather than a single
epoch.  For example, a deep monitoring survey also provides pre-explosion 
progenitor light curves to study variability and to search for signs of 
binarity (through eclipses). It would also permit the detection of SN obscured
by $A_V \simeq 10$~mag of extinction.

Spectacular outbursts such as that of $\eta$ Carina in the
19th century (Smith et al.~2003; Morse et al.~2001) and P Cygni
(Walborn 1976; Smith \& Hartigan 2006) around 1600 {\small CE} should
be recognizable by a characteristic brightening, reddening, and then
decay back to quiescence.  A number of extragalactic $\eta$
Carina-like outbursts have already been identified (the ``SN impostors'';
see Smith \& Owocki 2006 and references therein).  The observational
campaign required to identify failed SNe would be invaluable for identifying
lower-luminosity transients associated with such extreme mass loss
events and should measure their rates.  This could be extremely
important for understanding the role of such outbursts in the
late-time evolution and mass-loss of massive stars and, potentially,
for constraining the physics of their ignition.  The survey would
expand the number of supergiants for which order unity luminosity
variations would be detected by well over an order of magnitude (see
Fig.~\ref{fig:rate}).

Variability will be present in the massive star sample for a range of
other reasons.  Pulsations, whether periodic like Cepheids or more
irregular like luminous blue variables, are easy to recognize
given a reasonable number ($\sim 20$) of monitoring epochs.
Figure~\ref{fig:rate} illustrates that a full monitoring program for these
galaxies would be able to detect relatively weak 10\% luminosity
variations in $\sim 10^5$ supergiants.  Since $M_V\simeq -4$
corresponds to a 10~day Cepheid, essentially all Cepheids useful for
distance scale studies (see Macri et al. 2006) would be detected.
Eclipsing contact binaries can also be detected with
modest numbers of epochs (e.g., Prieto et al. 2008a), but significantly
detached systems would probably require a prohibitive number of epochs.
Other sources, such as RCB stars or novae,
correspond to fainter ``progenitors'' than the supergiants.
An $M_V=-3$ RCB star, if detected, would vanish and then
reappear, while a nova would appear and then disappear.  Both
phenomena are very different from (failed) SNe that start with a
luminous star, have a transient of some kind, and then ultimately have
no star.

\begin{figure}[t]
\plotone{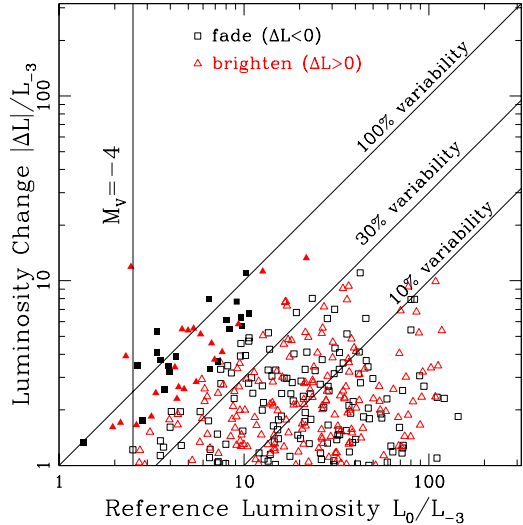}
\caption{Variable sources in M81.  The squares (triangles) show sources
as a function of their estimated luminosity $L_0$ in the reference image and
their decrease (increase) in luminosity $\Delta L$ between January and 
October 2007 in units of the luminosity $L_{-3}\simeq 1800 M_\odot$
corresponding to $M_V=-3$. 
The lines indicate the fractional variability, where a vanishing star
should lie on the 100\% variability line.  We inspected all sources with variability
$|\Delta L|/L_0 > 1/2$ and $L_0>L_{-3}$ (filled points) and found no 
candidate failed SNe. }
\label{fig:background}
\end{figure}

We explore the problem of backgrounds based on our 27 epochs of M81
monitoring data collected from 2007 January 16 to 2007 October 15, a
span of 272 days, using the Large Binocular Telescope (Prieto et al 2008a).
We characterize the sources by their V-band luminosity, $L_0$, in the reference 
image and the change in their luminosity, $\Delta L$, between the first and 
last epochs.  We will scale these by the luminosity $L_{-3}$ corresponding to 
$M_V=-3$ (about $1300L_\odot$ or 1100 counts in the reference image).     
We examine objects with large changes in their flux ($|\Delta L| > L_0/2$)
whose change in flux approaches that of a supergiant ($|\Delta L| > L_{-3}$,
see Fig.~\ref{fig:cmd} and \ref{fig:background}).  These criteria will also catch
high amplitude variability from objects that are not detected in the
reference image.  Of approximately 600 variable sources, only 53 meet
these criteria, of which 28 faded and 25 brightened.  Most of the candidates
are Cepheid (29) or other variable (19) stars for which the first and last epochs
coincidentally lie at maxima and minima of the light curves. Four sources appear
to be novae (2 bright in the first epoch and 2 bright in the last), 
and 1 is an artifact (diffraction spike).  All
but two of the sources are seen as discrete sources in the reference
image constructed from a stack of the 12 best epochs.  If we only look at the initial
and final epochs, where the initial epoch was taken in terrible 
conditions (FWHM 1\farcs8) and the final epoch in reasonable
conditions (FWHM 1\farcs0), 24 of the sources are seen in both
epochs, and 4 are seen in neither. For rising sources, 10 are
seen in the final but not the initial epoch, while for falling
sources one is seen in the initial but not the final epoch and 
two are seen in the final epoch but not the initial.  There were 
no plausible candidates for a failed SN.   We conclude that 
backgrounds are relatively easily controlled by combining sparse monitoring 
(to eliminate novae from the initial epochs and to detect the common 
large-amplitude variable stars) with direct inspection of the final
epoch.  This holds even if we restrict our analysis to using only
a modest fraction of the epochs available for M81. 
The steps for confirming a candidate should be to obtain 
additional epochs to further rule out other sources of variability
and to search for X-ray emission from the black hole.

\section{Discussion}
\label{sec:discuss}

While early reports that a previously known star was ``not to be found in 
the heavens'' (Cooper 1847; see also Herschel \& Flamsteed 1797) were 
probably due to issues in cataloging, it is now possible to systematically 
search for legitimate ``lost stars'' by direct monitoring.  Ultimately, this
will enable direct measurement of the rates of SNe and failed SNe
based on the properties of the evolved progenitors.  An 8-m telescope
with a wide field of view camera can simply watch enough supergiants
to detect any form of death, whether luminous or not.  At its
simplest, one subtracts the final image from the initial image and counts
the number of sources with supergiant luminosities that go missing.
This information, both temporal and spatial, can then trigger searches
for coincident bursts of neutrinos (Ando et al. 2005, Kowalski \& Mohr 2007) or
gravitational waves (Arnaud et al. 2004).
  In short, a modest investment of observing time over five years can
begin to measure and limit the fates of individual stars whether they
explode or simply collapse.  Moreover, the survey generates a wealth 
of new information on normal SNe, massive star evolution, binarity 
and the local distance scale as it proceeds.  Tests using monitoring
data for M81 from the Large Binocular Telescope (Prieto et al. 2008a)
found no false positives and formally set a limit that the rate of
failed SNe is $<80$ times that of normal SNe at 90\% confidence.    

Uncovering the existence of such ``unnovae'' would lead to a number of
interesting consequences, including significant changes in our
picture of metal enrichment and feedback.  
It would also change our
expectations for event rates in gravitational wave detectors, as
significant numbers of what are now expected to be NS-NS binaries
would instead be NS-BH or BH-BH binaries, which would both increase
the inspiral rates and change the expected signatures of coalescence
(Belczynski et al. 2007).    
The available observational evidence 
requires the most massive stars ($M\gtorder 25 M_\odot$) to become 
BHs and that many $8M_\odot \ltorder M \ltorder 25M_\odot$
stars become normal SN.  But theorists should also take seriously 
the implication from the difficulty of producing successful explosions 
that many of these less massive stars may also become BHs without a 
normal SNe.  Finally, while we argue that we can now detect BH
formation even if the signature is for a star to simply disappear,
it would be very helpful to have more quantitative estimates of the optical
signatures expected for the scenarios in Figure~\ref{fig:outcomes}. 
  
Using the new generation of telescopes to probe some of these issues
is not new.  There are several on-going programs to obtain the data
needed to characterize future SN progenitors in this volume
(e.g., Crockett et al. 2007).  There are trial programs that employ ground-based 
telescopes to do variability surveys using image subtraction at or near 
these depths focused on either microlensing (e.g., de Jong et al. 2007), Cepheids
and eclipsing binaries (e.g., Prieto et al. 2008a) or other variables
(e.g., Rejkuba et al. 2003).  What has not been
emphasized is that these surveys are on the verge of exploring a kind
of ``terra incognita'' where we can search for any phenomena occurring
at the rates of SNe but without the dramatic signatures of
SNe.  There is some potential for making similar studies with
historical data, if the extensive historical data on the Magellanic
Clouds, M31
and M33 can be combined to use the long time baselines (50-100
years) to compensate for the smaller number of stars.  In the future,
with the advent of large
scale synoptic surveys such as LSST (Tyson 2002), JDEM
(e.g., SNAP, Aldering et al. 2004) or EUCLID (e.g. DUNE, R\'efr\'egier et al.
2006), these studies will be very straightforward. 

We are reminded of the search for proton decay, a disappearance
campaign that has yet to succeed in its primary objective, yet
ultimately led to the discovery of ``new physics'' (massive
neutrinos), and demonstrated the feasibility of neutrino
astronomy (Hirata et al. 1987, Bionta et al. 1987).  
In regards to SNe, we have certainly been surprised
before, both with the blue supergiant progenitor of SN 1987A and the
non-detection of the Cas A supernova, to name just a few instances.
It would not be surprising, therefore, for such a survey
to unexpectedly discover ``new astrophysics''.

\acknowledgments

We thank Ethan Vishniac, Dieter Hartmann, Craig Wheeler and our
anonymous second referee for helping us through an unusual 
review process.  We would like to thank Roger Chevalier, Stuart Shapiro
and Craig Wheeler for their comments, as well as all the participants of the 
morning ``Astronomy Coffee'' at OSU, particularly Andrew Gould and Mark Pinsonneault.  
JLP and KZS acknowledge useful discussions with many participants of the ``Massive
Stars as Cosmic Engines" conference held in December 2007. This
research made use of the IAU Central Bureau for Astronomical Telegrams
and the Sternberg Astronomical Institute supernova catalogs and 
the NASA/IPAC Extragalactic Database (NED), which is operated by JPL/Caltech,
under contract with NASA.  JFB and HY are supported 
by NSF CAREER grant PHY-0547102,  MDK by DOE grant DE-FG02-91ER40690 and
JLP and KZS by NSF grant AST-0707982.


\begin{thebibliography}{99}
\frenchspacing



\bibitem[Alekseev \& Alekseyeva 2002]{}Alekseev, E.N., \& Alexeyeva, L.N., 2002, Soviet
  Journal of Experimental and Theoretical Physics, 95, 5

\bibitem[Aldering et al. (2004)]{astro-ph/0405232} Aldering, G., 
et al., 2004, astro-ph/0405232

\bibitem[Allard \& Lupton (1998)]{1998ApJ...503..325A} Alard, C. \&
Lupton, R. H., 1998, ApJ, 503, 325

\bibitem[Ando et al.(2005)]{2005PhRvL..95q1101A} Ando, S., Beacom,
J.F. \& Y{\"u}ksel, H., 2005, \prl, 95, 171101

\bibitem[Arnaud et al. (2004)]{}Arnaud, N., et al., 2004,
  Astroparticle Physics, 21, 201

\bibitem[Astier et al.(2006)]{2006A&A...447...31A} Astier, P., Guy,
J., Regnault, N., et al., 2006, A\&A, 447, 31

\bibitem[Balberg et al. (2000)]{2000ApJ...541...860B}Balberg, S., 
    Zampieri, L., \& Shapiro, S.L., 2000, ApJ, 541, 860

\bibitem[Balberg \& Shapiro (2001)]{2001ApJ...556...944}Balberg, S., 
    \& Shapiro, S.L., 2001, ApJ, 556, 944

\bibitem[Baumgarte et al. (1996)]{1996ApJ...468..823B}Baumgarte, T.W., Janka, H.-T.,
  Wolfgang, K., Shapiro, S.L., \& Teukolsky, S.A., 1996, 468, 823

\bibitem[Beacom et al. (2001)]{2001PhRvD..63g3011B} Beacom, J.F., Boyd, 
R.N., \& Mezzacappa, A., 2001, \prd, 63, 073011 

\bibitem[Belczynski et al. (2007)]{2007ApJ...662..504B}Belczynski, K., Taam, R.E., Kalogera, V.,
  Rasio, F., \& Bulik, T., 2007, ApJ, 662, 504

\bibitem[Belczynski \& Taam (2008)]{arXiv:0804.4143}Belczynski, K., \& Taam, R.,
  2008, ApJ submitted [arXiv:0804.4143]


\bibitem[Bethe et al., 2007]{2007PhR...442....5B} Bethe, H.A., Brown, G.E., \& Lee, C.-H.,
  2007, Physics Reports, 442, 5

\bibitem[Bionta et al., (1987)]{}Bionta, R.M.., et al., 1987, PRL, 58, 1494

\bibitem[Blondin et al.(2003)]{2003ApJ...584..971B} Blondin, J.M., 
Mezzacappa, A., \& DeMarino, C., 2003, ApJ, 584, 971 


\bibitem[Buras et al.(2003)]{2003PhRvL..90x1101B} Buras, R., Rampp, M., 
Janka, H.-T., \& Kifonidis, K., 2003, Physical Review Letters, 90, 241101 

\bibitem[Buras et al.(2006)]{2006A&A...457..281B} Buras, R., Janka, H.-T., 
Rampp, M., \& Kifonidis, K., 2006, \aap, 457, 281 



\bibitem[Burrows et al.(2006)]{2006ApJ...640..878B} Burrows, A., Livne, E., 
Dessart, L., Ott, C.D., \& Murphy, J., 2006, ApJ, 640, 878 


\bibitem[Cappellaro et al. (1999)]{1999AA...351..459} Cappellaro, E., Evans,
  R., \& Turatto, M., 1999, A\&A, 351, 459
  
\bibitem[Casares (2006)]{2006astro.ph.12312C}Casares, J., 2006, Black Holes: From Stars to
  Galaxies -- Across the Range of Masses, IAU Sympoisum 238 [astro-ph/0612312]

\bibitem[Chevalier (1993)]{1993ApJ...411L..33C}Chevalier, R.A., 1993, ApJL, 411, L33

\bibitem[Cooper (1847)]{1847MNRAS...8Q..16} Cooper 1847, \mnras, 8, 16

\bibitem[Crockett et al. (2007)]{2007MNRAS.381..835C}Crockett, R.M.,
   Smartt, S.J., Eldridge, J.J., Mattila, S., Young, D.R., Pastorello, A.,
   Maund, J.R., Benn, C.R., \& Skillen, I., 2007, MNRAS, 381, 835

\bibitem[Darnley et al. (2006)]{2006MNRAS.369..257D} Darnley, M.J.,
  Bode, M.F., Kerins, E., et al., 2006, MNRAS, 369, 257

\bibitem[de Jong et al. (2007)]{astro-ph} De Jong, J.T.A., Kuijken, K.H., \&
   H\'eraudeau, P., 2007, A\&A in press [astro-ph/0712.1052]

\bibitem[Doyle (1894)]{Doyle...1894} Doyle, A.C., 1892, Silver Blaze, in 
The Memoirs of Sherlock Holmes, (George Newnes: London)

\bibitem[Duez et al. (2004)]{2004PhRvD..69j4016D}Duez, M.D, Shapiro, S.L.
  \& Yo, H.-J., 2004, PhRvD, 69, 104016

\bibitem[Evans (1997)]{1997PASA...14..204} Evans, R 1997, PASA, 14, 204


\bibitem[Frieman et al. (2008)]{2008AJ....135..338F} Frieman, J. A.,
Bassett, B., Becker, A., et al. 2008, AJ, 135, 338

\bibitem[Fryer et al. (1996)]{1996ApJ...460..801F} Fryer, C.L., Benz, W., 
\& Herant, M., 1996, ApJ, 460, 801 

\bibitem[Fryer (1999)]{1999ApJ...522..413F} Fryer, C.L., 1999, ApJ, 522, 
413 

\bibitem[Fryer \& Warren (2002)]{2002ApJ...574L..65F} Fryer, C.L., \& 
Warren, M.S., 2002, ApJL, 574, L65 



 
\bibitem[Gould (2000)]{2000ApJ...535..928G} Gould, A., 2000, ApJ, 535, 928 

\bibitem[Gould \& Salim (2002)]{2002ApJ...572..944G} Gould, A., \& Salim, 
S., 2002, ApJ, 572, 944 

\bibitem[Hartman et al. (2006)]{2006MNRAS.371.1405H} Hartman, J. D.,
Bersier, D., Stanek, K. Z., et al. 2006, MNRAS, 371, 1405

\bibitem[Heger et al. (2003)]{2003ApJ...591..288H} Heger, A., Fryer, C.L., 
Woosley, S.E., Langer, N., \& Hartmann, D.H., 2003, ApJ, 591, 288

\bibitem[Herschel \& Flamsteed (1797)]{1797RSPT...87..293H} Herschel,
W. \& Flamsteed, J., 1797, RSPT, 87, 293

\bibitem[Hirata et al., (1987)]{}Hirata, K., et al., 1987, PRL, 58, 1490

\bibitem[Hopkins \& Beacom (2006)]{2006ApJ...651..142H}Hopkins, A.M., \& Beacom, J.F., 2006, ApJ, 651, 142

\bibitem[Ikeda et al. (2007)]{2007ApJ...669..519I}Ikeda, M., et al., 2007, ApJ, 669, 519



\bibitem[Karachentsev et al. (2004)]{2004AJ....127.2031K}
Karachentsev, I. D., Karachentseva, V. E., Huchtmeier, W. K., \&
Makarov, D. I. 2004, AJ, 127, 2031

\bibitem[Kowalski \& Mohr (2007)]{}Kowalski, M., \& Mohr, A., 2007, 
  Astropartical Physics, 27, 533

\bibitem[Lejeune \& Schaerer (2001)]{2001A&A...366..538L}Lejeune, T.,
\& Schaerer, D., 2001, A\&A, 366, 538

\bibitem[Li et al. (2000)]{2000AIPC..522..103L} Li, W. D., Filippenko,
A. V., Treffers, R. R., et al. 2000, AIPC, 522, 103

\bibitem[Li et al. (2007)]{2007ApJ...661.1013L} Li, W., Wang, X., Van
Dyk, S.D., et al. 2007, ApJ, 661, 1013

\bibitem[Liebend{\"o}rfer et al. (2001)]{2001PhRvD..63j3004L}
Liebend{\"o}rfer, M., Mezzacappa, A., Thielemann, F.-K., et al. 2001,
\prd, 63, 103004

\bibitem[Liebend{\"o}rfer et al. (2004)]{2004ApJS..150..263L}
Liebend{\"o}rfer, M., Messer, O.E.B., Mezzacappa, A., et al. 2004,
ApJS, 150, 263

\bibitem[Livne et al.(2004)]{2004ApJ...609..277L} Livne, E., Burrows, A.,
Walder, R., Lichtenstadt, I., \& Thompson, T.~A., 2004, ApJ, 609, 277 

\bibitem[MacFadyen \& Woosley (1999)]{1999ApJ...524..262M} MacFadyen, A.I., 
\& Woosley, S.E., 1999, ApJ, 524, 262 

\bibitem[Macri et al. (2006)]{2006ApJ...652.1133M} Macri, L.M.., Stanek, K.Z.,
  Bersier, D., Greenhill, L.J., \& Reid, M.J., 2006, AJ, 652, 1133


\bibitem[Massey (2002)]{2002ApJS..141...81M} Massey, P., 2002, ApJS, 141, 81

\bibitem[Miknaitis et al. (2007)]{2007ApJ...666..674M} Miknaitis, G.,
Pignata, G., Rest, A., et al. 2007, ApJ, 666, 674


\bibitem[Morse et al.(2001)]{2001ApJ...548L.207M} Morse, J.A., Kellogg,
J.R., Bally, J., Davidson, K., Balick, B., \& Ebbets, D., 2001, ApJL,
548, L207 

\bibitem[Muno et al. (2006)]{2006ApJ...636L..41M} Muno, M.P.,. Clark, J.S., Crowther, P.A., et al., 2006, ApJL, 636, L41


\bibitem[Ohnishi et al.(2006)]{2006ApJ...641.1018O} Ohnishi, N., Kotake,
K., \& Yamada, S., 2006, ApJ, 641, 1018 

\bibitem[Orosz et al. (2007)]{2007Natur.449..872O} Orosz, J. A.,
McClintock, J.E., Narayan, R., et al. 2007, Nature, 449, 872

\bibitem[Prieto et al. (2008a)]{2008ApJ...673L..59} Prieto, J. L.,
Stanek, K. Z., Kochanek, C. S., et al. 2008a, ApJ, 673, L59

\bibitem[Prieto et al. (2008b)]{arXiv:0803.0324} Prieto, J. L.,
Kistler, M.D., Thompson, T.A., et al., 2008b, ApJL submitted [arXiv:0803.0324]

\bibitem[Rampp \& Janka (2000)]{2000ApJ...539L..33R} Rampp, M., \& Janka, 
H.-T., 2000, ApJL, 539, L33 

\bibitem[R\'efr\'egier et al. (2006)]{2006SPIE.6265E..58R}
R\'efr\'egier et al. (2006) et al. 2006, SPIE, 6265, 62651

\bibitem[Rejkuba et al. (2003)]{2003A&A...406...75}Rejkuba, M., 
  Minniti, D., \& Silva, D.R., 2003, A\&A, 406, 75

\bibitem[Riess et al. (2004)]{2004ApJ...607..665R} Riess, A. G.,
Strolger, L-G., Tonry, J., et al. 2004, ApJ, 607, 665

\bibitem[Scheck et al. (2008)]{2008A&A...477..931S} Scheck, L., Janka, 
H.-T., Foglizzo, T., \& Kifonidis, K., 2008, \aap, 477, 931 


\bibitem[Smartt et al. (2004)]{2004Sci...303..499S} Smartt, S.J.,
  Maund, J.R., Hendry, M.A., Tout, C.A., Gilmore, G.F., Mattila, S.,
  \& Benn, C.R., 2004, Science, 303, 499

\bibitem[Smith et al.(2003)]{2003AJ....125.1458S} Smith, N., Gehrz, R.D.,
Hinz, P.M., Hoffmann, W.F., Hora, J.L., Mamajek, E.E., \& Meyer, M.R., 2003, AJ, 125, 1458 

\bibitem[Smith \& Hartigan(2006)]{2006ApJ...638.1045S} Smith, N., \&
Hartigan, P., 2006, ApJ, 638, 1045 

\bibitem[Smith \& Owocki(2006)]{2006ApJ...645L..45S} Smith, N., \& Owocki,
S.P., 2006, ApJL, 645, L45 

\bibitem[Stanek et al. (2006)]{2006AcA....56..333S}Stanek, K.Z.,
  Gnedin, O.Y. Beacom, J.F., Gould, A.P., Johnson, J.A.,
  Kollmeier, J.A., Modjaz, M., Pinsonneault, M.H., Pogge, R.,
  \& Weinberg, D.H., 2006, AcA, 56, 333

\bibitem[Sumiyoshi et al.(2005)]{2005ApJ...629..922S} Sumiyoshi, K.,
Yamada, S., Suzuki, H., Shen, H., Chiba, S., \& Toki, H., 2005, ApJ, 629,
922 

\bibitem[Sumiyoshi et al. (2007)]{2007ApJ...667..382S} Sumiyoshi, K., 
Yamada, S., \& Suzuki, H., 2007, ApJ, 667, 382 

\bibitem[Thompson et al. (2003)]{2003ApJ...592..434T} Thompson, T.A., 
Burrows, A., \& Pinto, P.A., 2003, ApJ, 592, 434 


\bibitem[Tyson (2002)]{2002SPIE.4836...10T} Tyson, J.A., 2002, \procspie, 
4836, 10 


\bibitem[van den Bergh, Sidney \& Tammann (1992)]{1991ARA&A..29..363V}
van den Bergh, S., \& Tammann, G. 1991, ARA\&A, 29, 36

\bibitem[Walborn(1976)]{1976ApJ...204L..17W} Walborn, N.R., 1976, ApJL, 204, L17 



\bibitem[Woosley \& Weaver (1995)]{1995ApJS..101..181W} Woosley, S.E., \& 
Weaver, T.A., 1995, ApJS, 101, 181 

\bibitem[Zampieri et al. (1998)]{1998ApJ...505..876Z} Zampieri, L., Colpi, M.,
  Shapiro, S.L., \& Wasserman, I., 1998, ApJ, 505, 876

\bibitem[Zhang et al. (2007)]{astro-ph/0701083}Zhang, W., Woosley,
S.E., \& Heger, A., 2007, ApJ in press [astro-ph/0701083]

\end{thebibliography}
\end{document}